\documentstyle[prl,multicol,twoside,epsf,epsfig,,aps]{revtex}

\title{Quantum decay rates for driven barrier potentials in the
strong friction limit} 

\author{Joachim Ankerhold}
\address{
Fakult{\"a}t f{\"u}r Physik, Albert-Ludwigs-Universit{\"a}t
Freiburg,
 Hermann-Herder-Stra{\ss}e 3, D-79104 Freiburg, Germany}

\date{Rapid Communication, Phys. Rev. E, in press}

\begin{document}

\maketitle

\begin{abstract}
Quantum decay rates for barrier potentials driven by external
stochastic and periodic forces in the strong damping regime are
studied. Based on the recently derived quantum Smoluchowski equation
[Phys. Rev. Lett. {\bf 87}, 086802 (2001)] explicit analytical and numerical
results are presented for the case of the resonant activation
phenomenon in a bistable potential and the escape from a metastable
well with oscillating barrier, respectively. The
significant impact of quantum 
fluctuations  is revealed.
\end{abstract}
\vspace{-0.6cm}
\hspace{1cm}{\small PACS:05.40-a,82.20.Pm,03.65.Yz}

\vspace*{-0.2cm}
\begin{multicols}{2}
{\it Introduction--} Escape over a high potential
barrier driven by thermal noise is of fundamental interest in physics
and chemistry \cite{kramers}. And it is meanwhile 
well understood: after a transient period of time the
decay is governed by a rate 
constant, which in the simplest classical case is of  Arrhenius
type provided the decaying system stays in thermal
equilibrium far from the barrier. 
 What happens if due to
additional external forces the system is far from thermal equilibrium? This
question has recently gained much attention for classical barrier
transport in the strong damping limit (Smoluchowski limit). 
Important examples are stochastic resonance \cite{jung}, resonant
activation in biochemical 
reactions and tunnel diodes \cite{fluc1d}, or directed motion in
ratchet systems \cite{astumian}.  
While even the classical physics of such phenomena is in many  cases
not yet completely explored, much less is known about
corresponding quantum systems.

Namely, in contrast to the classical
range, tractable equations of motions of quantum
dissipative systems also for strong damping and low temperatures do
not exist \cite{weiss}. The path integral representation  provides an exact
expression for the time dependent density matrix, but even a numerical
evaluation is in general prohibitive. Particular progress has been
made with the 
development of Master equations \cite{redfield} and the 
quasiadiabtic propagator approach \cite{makri}. This way, 
extensive studies exist for bistable
systems driven by external periodic forces \cite{qsr}. However, these
and related techniques 
require--either to be valid or to be practicable--that energy
level broadening due to friction remains sufficiently small (depending
on the approach) so that
for the relevant dynamics  
the Hilbert space of the bare system can be reduced to a few
lowest lying eigenstates. Of course, this condition fails
in the domain of very strong friction.  
Recently we showed \cite{prl}
that exactly in this range crucial simplifications arise:   
As in the classical Smoluchowski limit, momentum equilibrates on time
scales much faster than any other time scales; accordingly, an
equation of motion--the so-called quantum Smoluchowski equation--can be
derived for the
position probability distribution from the exact path integral
result. The influence of 
quantum fluctuations turns out to be substantial. Hence, for the first
time we are now in a position to explore driven 
barrier escape for arbitrary overdamped quantum systems \cite{remark}.

Here, we focus on two paradigmatic examples, namely, a bistable
potential with a barrier fluctuating randomly in time and a metastable
well driven by a periodic monochromatic force. Classically, for the first
case the phenomenon of resonant activation is
characteristic, while for the second one a substantial rate enhancement
by driving is observed. We elucidate the significant impact of  quantum
fluctuations  on both processes.

{\it Quantum Smoluchowski--}For a systems coupled to a heat bath
environment the reduced density matrix follows from $\rho(t)={\rm
Tr}_b\{W(t)\}$ where Tr$_b\{\cdot\}$ denotes the trace over bath degrees
of freedom and $W(t)$ is the time dependent density matrix of
the system+bath compound. Within the path integral
approach a formally exact expression for the position representation
$\rho(q,q',t)$ can be given \cite{weiss}.
 Now, in \cite{prl} it was shown
that within 
the  Quantum-Smoluchowski range (QSR), i.e.\ 
\begin{equation}
 \gamma/\omega_0^2\gg\hbar\beta, 1/\gamma\ \ \ \ \mbox{ and} \ \ \ \
\hbar\gamma\gg 
k_{\rm B} T, \label{aq1}
\end{equation}
the position distribution $P(q,t)=\rho(q,q,t)$ is well determined by a
simple time evolution equation coined
Quantum Smoluchowski equation (QSE).
Here $\gamma$ denotes the friction constant, $\omega_0$ the ground
state frequency in a potential $V(q)$, and $\beta=1/k_{\rm B} T$
inverse temperature. This limit is opposite to the classical Smoluchowski
range where 
$\omega_0\hbar\beta\ll 1$ and $k_{\rm B} T\gg \hbar\gamma$.
The QSE reads
$\dot{P}=(1/M\gamma)\partial_q \hat{L}_{\rm qm} P$ where
\begin{equation}
\hat{L}_{\rm qm}=V'+\lambda V'''/2+k_{\rm
B}T\partial_q [1+\lambda\beta V''] \label{qmsmolu}
\end{equation}
with $V'=d V(q)/dq$. Further, 
\begin{equation}
\lambda=
({\hbar}/{\pi
M\gamma})\, \ln(\hbar\beta\gamma/2\pi)\label{lambda}
\end{equation}
accounts for the dominating impact of  quantum fluctuations.
Equivalently, the dynamics of an overdamped quantum systems can be seen as
 a 
classical Smoluchowski dynamics with an effective
potential $V_{\rm eff}=V+\lambda/2 V''$ and an effective diffusion term
$D_{\rm eff}= k_{\rm B} T 
(1+\lambda\beta V'')$. Note that quantum fluctuations in
Eq.~(\ref{qmsmolu}) are of order $\ln(\gamma)/\gamma$ and thus, are
much larger than classical finite friction corrections which are of
order $1/\gamma^2$.

{\it Static barriers--}We first recall results for
the decay rate in 
static barrier potentials. There, adjacent to the barrier at $q=q_b$   we assume a well
region around a minimum at $q=q_0$ such that
the barrier height $V_b$ obeys $V_b\gg k_{\rm B}
T,\hbar\omega_0$. Moreover, we take  smooth potentials for
granted. Then, in \cite{prl} it was 
shown that the escape rate out of the well is given within the QSR as
\begin{equation}
\Gamma_{QSR}= \frac{\sqrt{V''(q_0) |V''(q_b)|}}{M\gamma} {\rm
e}^{-\beta V_b}\, {\rm e}^{\lambda\beta
[V''(q_0)+|V''(q_b)|]}.\label{qserate}
\end{equation}
The substantial rate increase due to
quantum fluctuations well agrees with the exact result
\cite{prl} (in the limit $\beta V_b\gg 1$).
Let us now turn to barrier potentials driven by external
forces. 

{\it Fluctuating barriers--}Thermally activated diffusion over a
potential barrier that fluctuates randomly in time has evoked much interest
recently. In the classical Smoluchowski limit it was shown that the 
interplay between relaxation, by thermally activated barrier passage,
and fluctuation, by correlated external noise, leads to a strongly
enhanced reaction rate in the resonant activation regime
\cite{fluc1d}. Here, we 
present the first study to this phenomena for corresponding quantum
systems. We look at the process of dichotomous barrier fluctuations
with a rate $\eta$ in a
symmetric double well and search for the ultimate
decay rate $k(\eta)$ of 
relaxation to equilibrium. Since for this  problem an analytical
classical theory was derived \cite{physica}, we can adapt the general
technique. In particular, in the  case of high barriers
considered here, it was shown that rates from the analytical theory
are in excellent agreement with numerically exact results
\cite{physica}. 

If the potential flips randomly between two surfaces $V_+$ and
$V_-$ at a rate $\eta$, we need two probability densities $P_+(q,t)$
and $P_-(q,t)$ with $P_+$ [$P_-$] being the density to find a particle at
time $t$ at position $q$ and the potential in state $V_+$
[$V_-$]. Accordingly, the 
two-dimensional QSE reads $\partial_t \vec{\rho}=S_\eta \vec{\rho}$ with
$\vec{\rho}=(P_+,P_-)$ and 
\begin{equation}
S_\eta=\left(\begin{array}{cc}
\hat{L}_+-\eta &\eta\\
\eta &\hat{L}_--\eta
\end{array}\right).\label{fluc1e}
\end{equation}
Here, $\hat{L}_\pm=(1/M\gamma)\partial_q \hat{L}_{\rm qm}^{(\pm)}$ with
potentials
$V_\pm=U\pm g$. The function $g$ describes the barrier modulations and is
assumed to have the following properties: it is symmetric around the
barrier top at $q=q_b$ and monotone decreasing away from it; outside
some finite range around the top it is zero. In particular, $V_+=V_-$
around the well minima located at $q=\pm q_0$. Further, its maximum
$g(q_b)$ is small compared to the barrier height $U_b$ but not
necessarily small compared to $k_{\rm B} T$. The ultimate decay rate
$k(\eta)$ is now defined as the least negative eigenvalue to the
operator $S_\eta$.
\vspace*{-3.5cm}
\begin{figure}
\vspace{-1cm}
\center
\epsfig{file=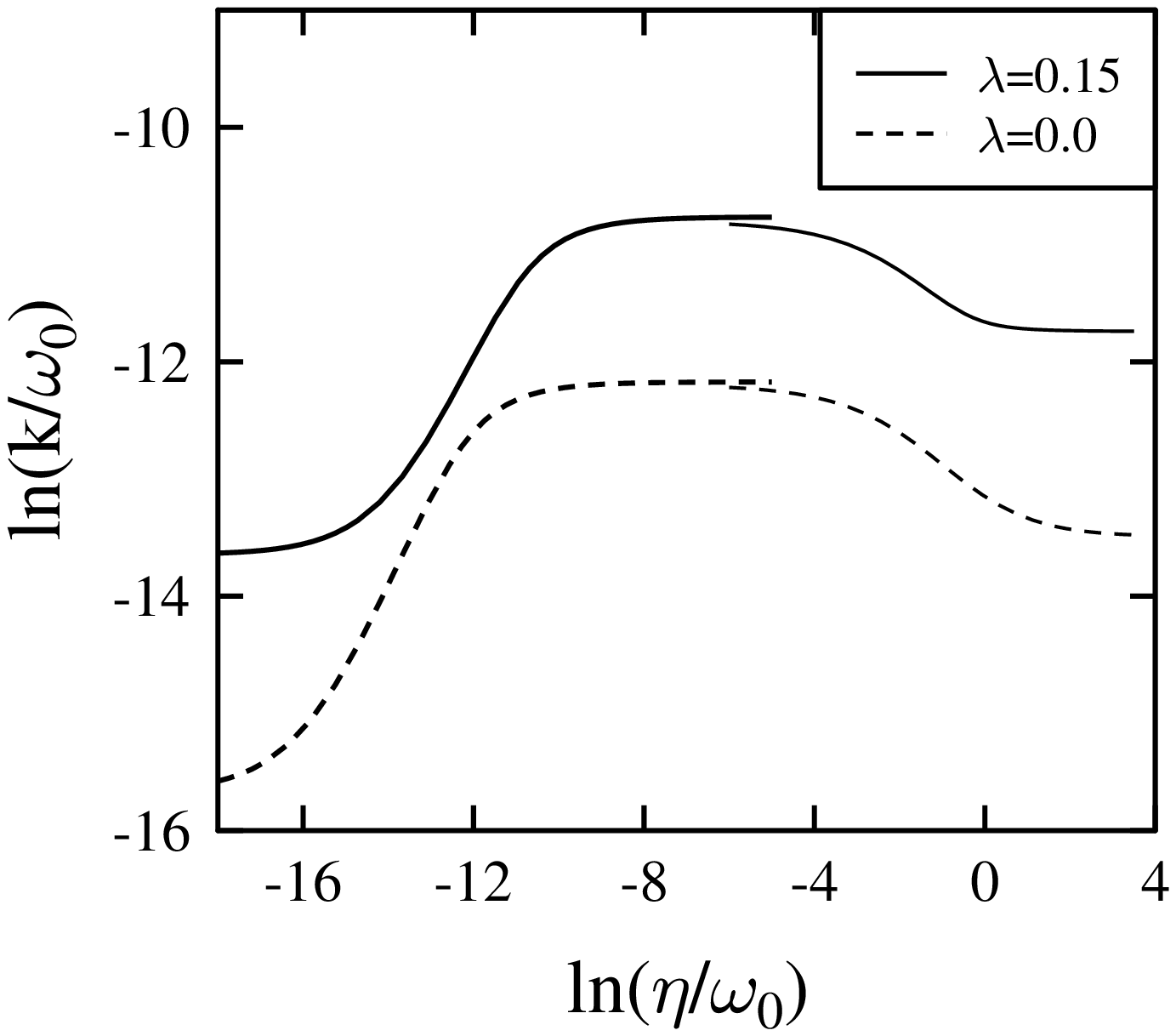,width=9cm}
\vspace{-4.2cm}
\label{ankerhold1}
\end{figure}
\noindent
\parbox{8.5cm}{\small{FIG.~1 Quantum (solid) and
classical (dashed) 
decay rate vs. the barrier fluctuation rate in a double well potential
with height $U_b/\hbar\omega_0=4$ and  $\omega_0\hbar\beta=3$. Shown
are the adiabatic approximation Eq.~(\ref{adia}) (thick) and the
effective potential approach Eq.~(\ref{flucrate}) (thin).}}
\vspace{0.2cm}

In the
adiabatic range of  small
$\eta$, qualitatively, there is no much change to the classical
result. After
each flip the system has enough time to relax to equilibrium inside
the well regions of the instantaneous potentials. Consequently, the
rate follows as the least negative eigenvalue of $S_\eta$ when
replacing the operators $\hat{L}_{\rm qm}^{(\pm)}$ by the static
quantum rates   
$-\Gamma_\pm$ for the individual potentials, see Eq.~(\ref{qserate}).
Hence, similarly to the classical rate we obtain
\begin{equation}
k(\eta)=(\Gamma_++\Gamma_-)/2+\eta-
\left[\eta^2+(\Gamma_+-\Gamma_-)^2/4\right]^{1/2}. \label{adia}
\end{equation}
As expected $k(0)=\Gamma_+$ and $k(\eta)\to k_{\rm res}\equiv
(\Gamma_++\Gamma_-)/2$ for 
$\eta\to \infty$. 

More interesting is the region 
 of moderate to fast barrier flippings,
i.e. $\eta\gg k(\eta)$.
There, the procedure is to solve $S_\eta \vec{\rho}\approx 0$ with the
boundary condition $\vec{\rho}(q_b)=0$ so that due to symmetry 
 $\vec{\rho}(q)=-\vec{\rho}(-q)$.
 This way, we solve the equation for
 the equilibrium eigenfunction, i.e.\ with zero eigenvalue, but under
 boundary conditions 
 corresponding to the
 relaxation eigenfunction with the least negative eigenvalue. As long
 as $k(\eta)$ is the smallest frequency in the system this
 approximation is justified. 
Now, in the
limit of a high barrier we employ a semiclassical type of 
ansatz $\vec{\rho}= \vec{z} \exp(-\phi/k_{\rm B} T)$ with an effective
potential $\phi$ and a prefactor accounting for 
terms of higher order in $k_{\rm B} T$. To obtain the dominant
 exponential contribution we do the following: 
The two coupled second order
 differential equations corresponding to Eq.~(\ref{fluc1e}) are
 transformed to four coupled equations of first order. This linear
 system is solved by diagonalizing the coefficient matrix. The
 relevant eigenvalue turns out to be $-\phi'$ and is determined by the
 solution to the cubic equation
\begin{equation}
2k_{\rm B}T \eta (1-\lambda\beta
{U}'')(\tilde{U}'-\phi')=\phi'(\tilde{V}'_+-\phi')
(\tilde{V}'_--\phi')\label{cubic} 
\end{equation}
obeying $0\leq \phi' \tilde{U}'\leq (\tilde{U}')^2$. Here, we
introduced effective 
potentials which are related to the bare potentials by
$\tilde{Y}=Y+(3\lambda/2) Y''-(\lambda\beta/2)(Y')^2$ for 
$Y=U,V_+,$ and $V_-$.  Of course, by formally putting $\lambda=0$ in
(\ref{cubic}) one regains the known classical result \cite{physica}. 
For the prefactor we  solve perturbatively the differential
equations $S_\eta\vec{\rho}=0$. Since this scheme works
similarly as in the classical case we refer to the
literature for further details \cite{physica}. 
  
For the rate one first derives from $S_\eta\vec{\rho}=-k\vec{\rho}$
that $k(\eta)\approx[P_+'(q_b)+P_-'(q_b)]/\{M\gamma\beta\int
[P_++P_-]\}$. Then, inserting the results for $P_\pm$ we gain
\begin{equation}
k(\eta)= [\Omega(\eta) \omega_0/\gamma]\, {\rm e}^{-\beta
\phi(q_b)}\, {\rm e}^{\lambda\beta
[U''(q_0)-|\phi''(q_b)|/2]}\label{flucrate}
\end{equation}
with $\omega_0^2=U''(q_0)/M$. The frequency $\Omega(\eta)$ coincides with the 
classical result \cite{physica} and thus, its lengthy expression is omitted
here. However, quantum fluctuations strongly affect the 
dominant exponential factor. In the
limits of very slow and very fast barrier fluctuations
$\lambda$-dependent terms lead basically to a renormalization of
temperature. For small $\eta$ we derive from Eqs.~(\ref{flucrate}) and
(\ref{fluc1e}) that $k(\eta)\to \Gamma_+$. In the region of motional
narrowing $\eta\to \infty$ it is $k(\eta)\to \Gamma_U$ where $\Gamma_U$
is given by Eq.~(\ref{qserate}) with $V$ replaced by the average
potential $U$.
\vspace*{-1.95cm}
\begin{figure}
\center
\epsfig{file=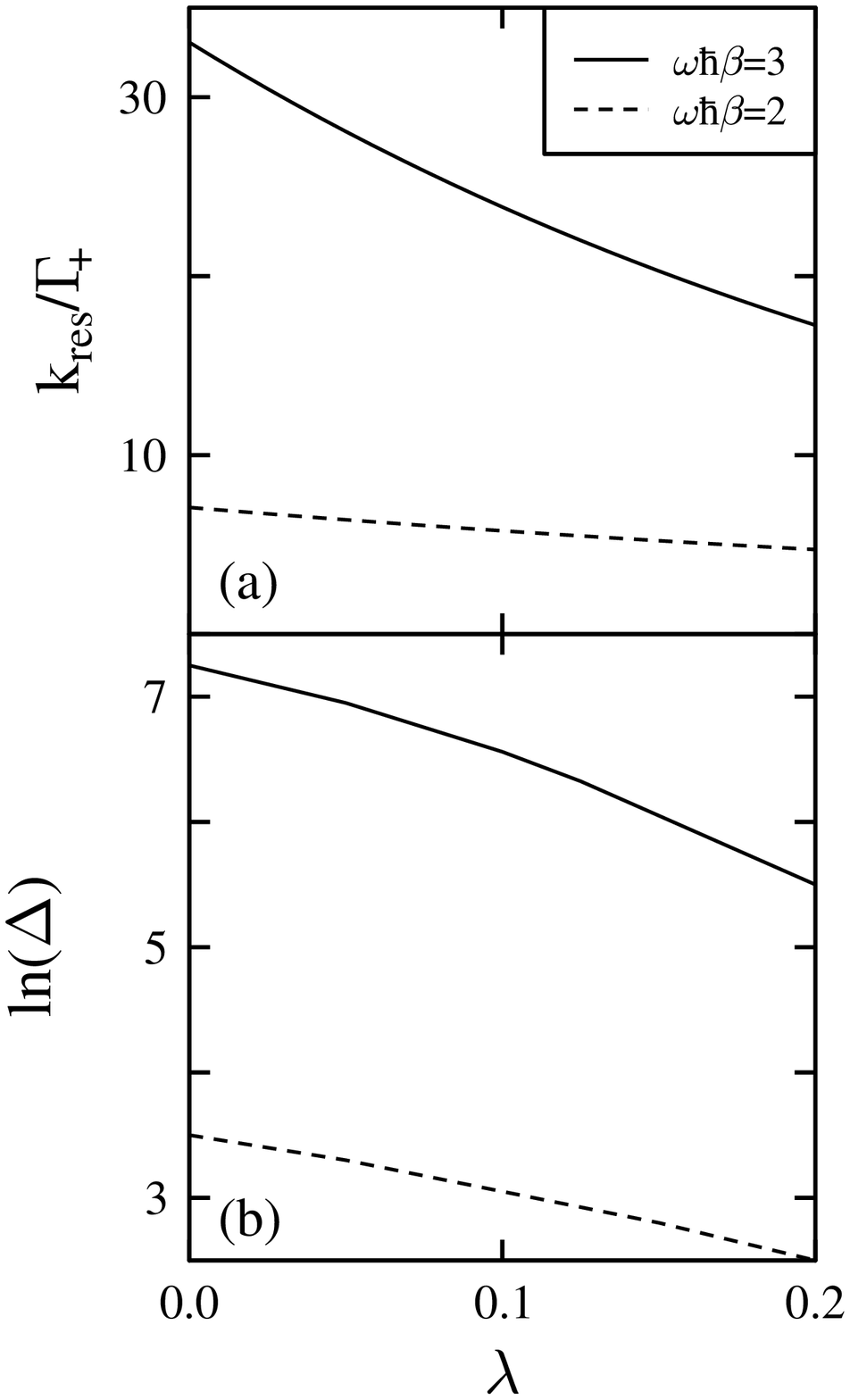,width=9cm}
\vspace{-3.2cm}
\label{ankerhold2}
\end{figure}
\noindent
\parbox{8.5cm}{\small{FIG.~2 (a) Ratio of the resonant activation rate
$k_{\rm res}$ to 
$k(0)=\Gamma_+$ and (b)  width of the resonant
activation maximum  vs. $\lambda$ for two different
temperatures. Other parameters  as in Fig.~1.}}
\vspace{0.2cm}

To study the ultimate quantum decay rate for all values of $\eta$ we evaluated
Eq.~(\ref{adia}) and Eq.~(\ref{flucrate}) with Eq.~(\ref{cubic})
numerically.
Results are displayed in Figs.~1, 2.
Besides an obvious rate increase, the major effects of
quantum fluctuations are to {\em substantially decrease} the relative
height as well as the width  of the resonant activation maximum. 
The shrinking relative height $k_{\rm
res}/\Gamma_+=(\Gamma_++\Gamma_-)/2\Gamma_+$ is a consequence of
effectively reduced barrier modulations 
$g(q_b)\to g(q_b)[1-\lambda |g''(q_b)|]$. The decreasing width of the
plateau range is ascribed for sufficiently large $\eta$
 to the fact that
then  quantum fluctuations on the
left hand side in Eq.~(\ref{cubic}) lead to an effectively enhanced
flipping rate $\eta\to  
\eta [1+\lambda \beta |U''(q_b)|]$ near the barrier top.  For
intermediate fluctuation rates the cubic equation (\ref{cubic}) gives
rise to a nonlinear dependence 
on $\lambda$ leading to an intimate relation between external barrier
fluctuations and intrinsic quantum fluctuations.

{\it Oscillating barriers--}A prominent example for a system far from
equilibrium is a metastable potential driven periodically by an external
force. Recently, in the classical Smoluchowski limit, decay rates were
studied for weak and moderate to strong driving as well
\cite{osci,lehmann,stein}. Here, we 
explore the corresponding quantum problem by calculating
numerically the time dependent decay rate $\Gamma(t)$ from
Eq.~(\ref{qmsmolu}) for long times. The time dependent potential is
chosen as $V(q,t;a)= V_0(q)+q  a\sin(\Omega t)$
with a static barrier $V_0(q)=-(M\omega_b^2/2) q^2[1+q/q_0]$. The
static barrier top is located at $q=0$, the well 
minimum at $q=-2 q_0/3$, and the static  height is $V_b/M\omega_b^2=10
q_0^2/27$. For the driven case ($a\neq 0$) the location of the
barrier top moves periodically in time
with $q_b(t)\approx (a/M\omega_b^2) \sin(\Omega t)$.
Consequently, 
 as in the classical case,  the rate
$\Gamma(t)=-\dot{N}(t)/N$
with $N(t)=\int_{-\infty}^{q_b(t)}P(q,t)$ also oscillates in time. 
Then, from
$\dot{P}=(1/M\gamma)\partial_q \hat{L}_{\rm qm} P$  we derive up to
negligible corrections 
\begin{equation}
\Gamma(t)= - \frac{k_{\rm B} T}{M\gamma}(1+\lambda\beta
V''_0[q_b(t)])\frac{\partial_q P[q_b(t),t]}{N(t)}.\label{oscirate}
\end{equation}
Note that for high barriers and times $t\ll 1/\Gamma(t)$ one can put
 $N(t)\approx 1$ if initially one starts from a normalized
 distribution $P(q,0)$.
\vspace*{-3cm}
\begin{figure}
\center
\epsfig{file=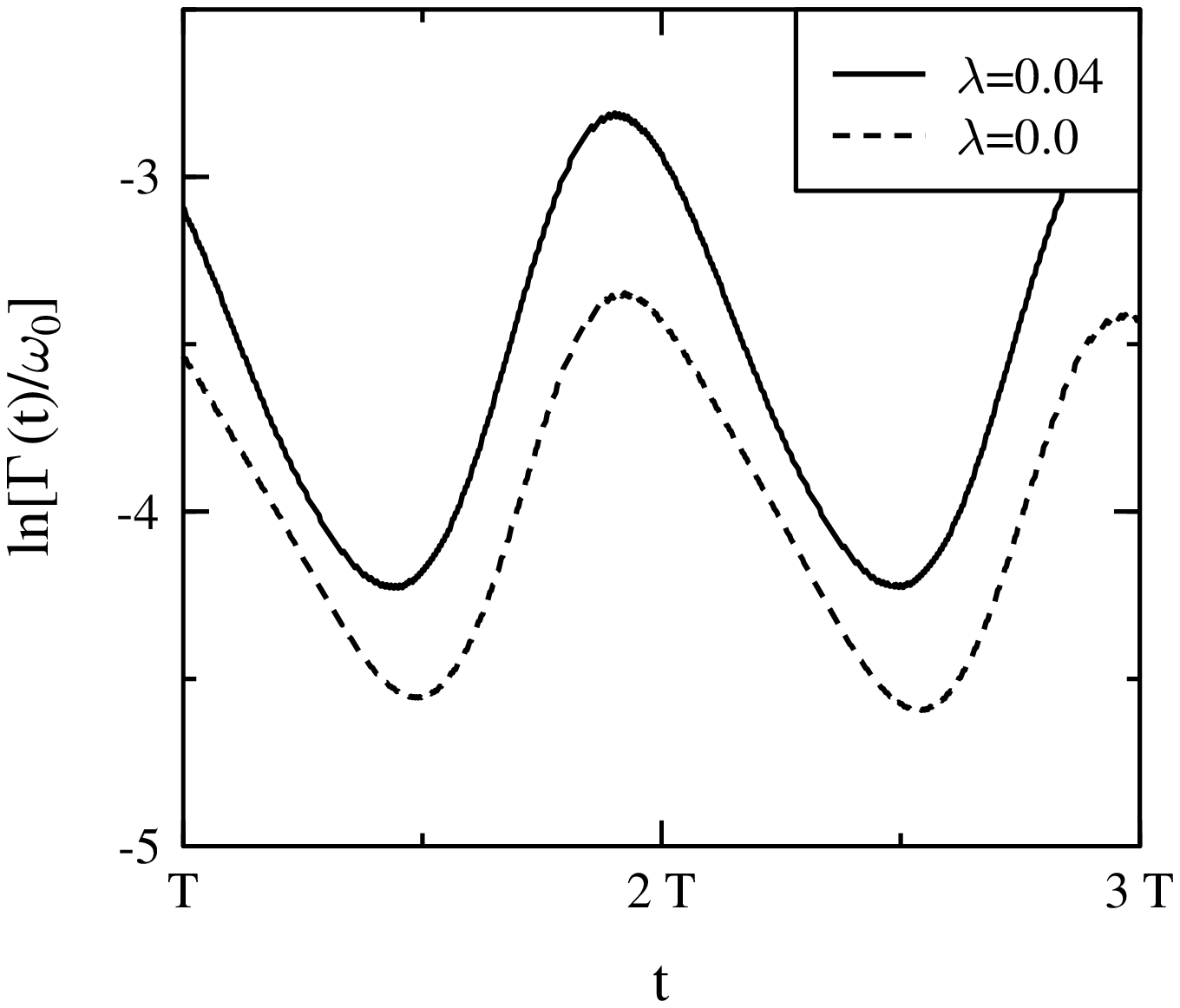,width=8cm}
\vspace{-4.cm}
\label{ankerhold3}
\end{figure}
\noindent
\parbox{8.5cm}{\small{FIG.~3 Time dependent quantum (solid) and
classical (dashed) rate $\Gamma(t)$ for $\Omega/\omega_b=2\pi$
($T=2\pi/\Omega$), 
$A=a/\sqrt{\hbar M\omega_b^3}=0.5$, $\omega_b\hbar\beta=3$, and
$V_b/\hbar\omega_b=4/3$.}}
\vspace{0.2cm}

High precision numerical results based on the QSE
[Eq.~(\ref{qmsmolu})] are shown in Fig.~3 for the 
time dependent rate $\Gamma(t)$. Apart from a rate enhancement the
influence of quantum
fluctuation is two-fold: first, $\Gamma(t)$ tends to be more
symmetric around its minima and maxima and second, both extrema are
slightly shifted to the left. In \cite{lehmann} Lehmann et
al.\ developed a nice theory for the pure 
classical case ($\lambda=0$) based on an asymptotic expansion for
$\beta V_b\gg 1$ (for a related theory see also
\cite{stein}). There, one finds a weak dependence of the 
location of the extrema on temperature. Since  locally in position
space the $\lambda$ dependence can be partially  incorporated in an effective
temperature, we  
attribute the observed shift in the quantum case to a similar kind of
process. However, what is not possible is to mimic the
numerical data for finite $\lambda$ by an effective temperature only.
\vspace*{-3cm}
\begin{figure}
\center
\epsfig{file=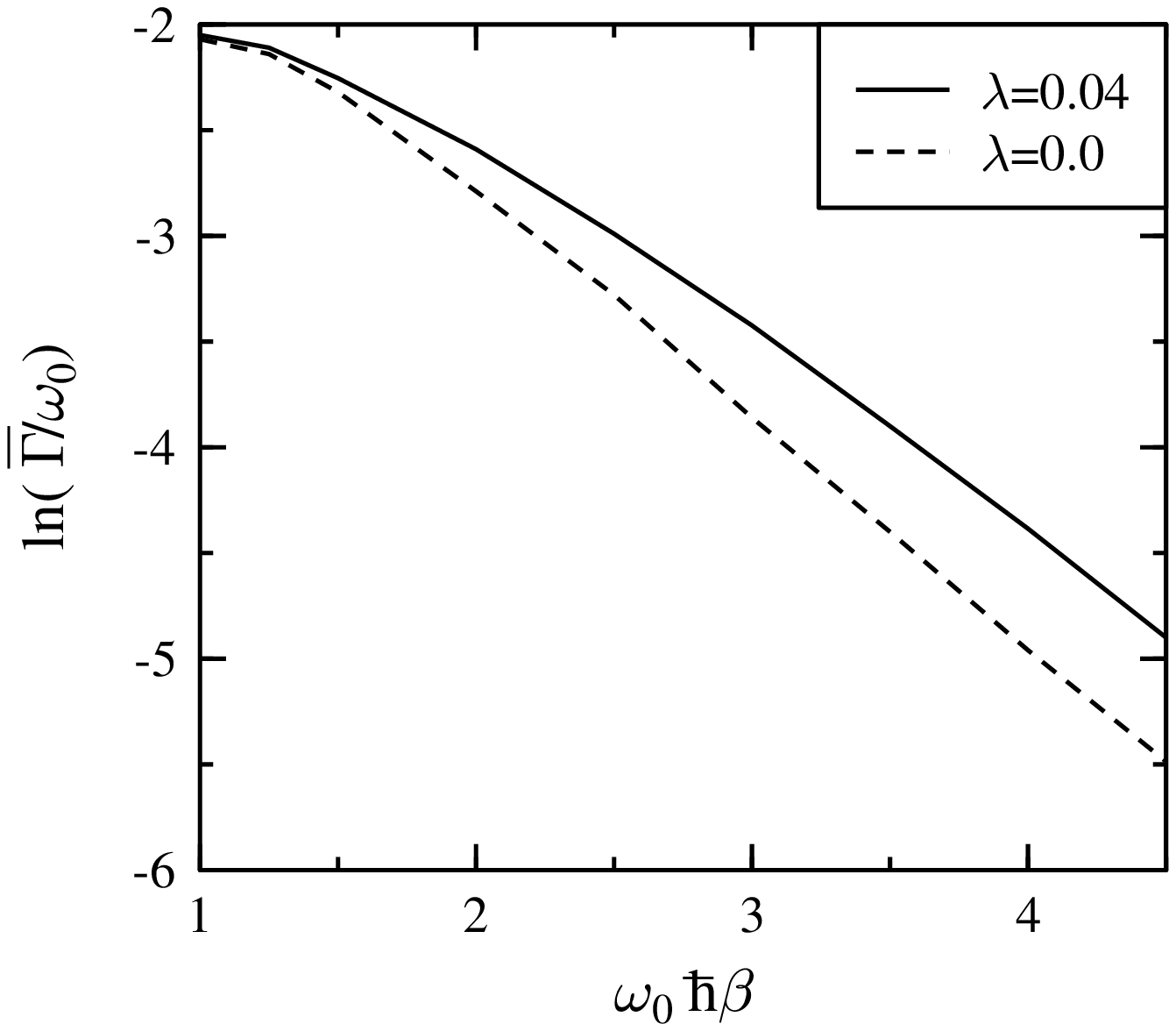,width=8cm}
\vspace{-4cm}
\label{ankerhold4}
\end{figure}
\noindent
\parbox{8.5cm}{\small{FIG.~4 Arrhenius plot of the time-averaged
quantum (solid) and 
classical (dashed) rate $\bar{\Gamma}$. Other parameters as  in
Fig.~3.}}
\vspace{0.2cm}

The averaged rate
\begin{equation}
\bar{\Gamma}=\lim_{\omega_b t> 1}\frac{1}{T}\int_t^{t+T}dt'\
\Gamma(t')\ ,\ \ T=\frac{2\pi}{\Omega}\label{avrat}
\end{equation}
 as a function either of inverse temperature or
driving amplitude is depicted in Figs.~4, 5. Again 
the rate enhancement is significant. Interestingly, while on a logarithmic
scale  the classical rate shows  a simple linear behavior as a function of 
$\omega_0\hbar\beta$  for sufficiently low temperatures (according to
the well-known exponential Arrhenius law), the quantum rate exhibits a
weak nonlinearity. Of 
course, in the high temperature limit quantum corrections become
negligible. As a function of the driving amplitude the difference
between quantum and classical rate shrinks with increasing
amplitude. As for strong driving most of the escape happens to occur when the
barrier is low and thus, is much less effective in hindering the transport,
the effect of a $\lambda$ induced rate enhancement due to a smaller
barrier height in $V_{\rm eff}=V+\lambda/2 \, V''$ is diminished.
\vspace*{-3cm}
\begin{figure}
\center
\epsfig{file=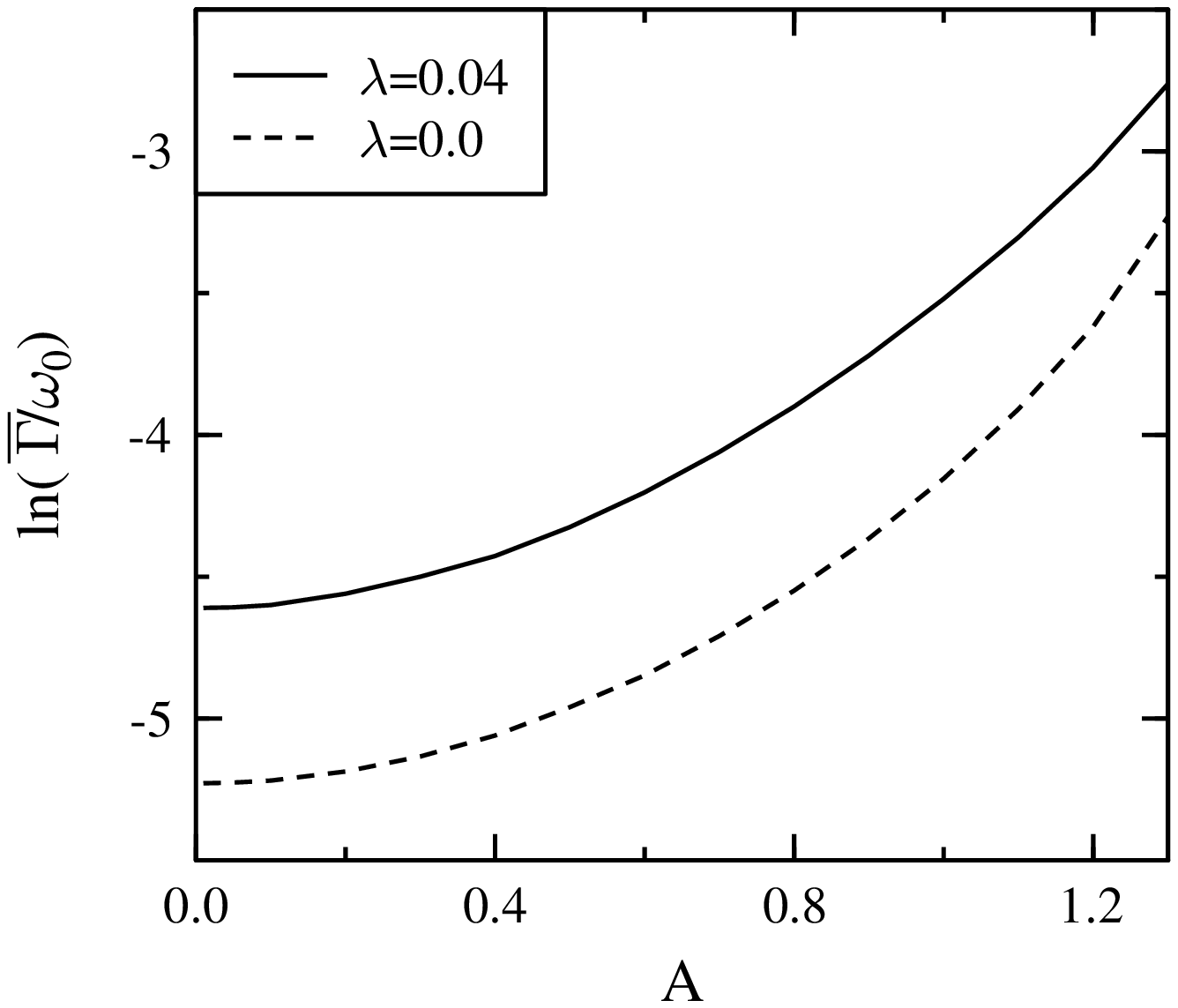,width=8cm}
\vspace{-4cm}
\label{ankerhold5}
\end{figure}
\noindent
\parbox{8.5cm}{\small{FIG.~5 Time-averaged quantum (solid) and
classical (dashed) rate as 
a function of the driving amplitude 
$A=a/\sqrt{\hbar M\omega_b^3}$. Other parameters as in
Fig.~3.}}
\vspace{0.2cm}

What about an analytical theory? We
think that in principle the classical theory \cite{lehmann,stein} can be
generalized to the quantum case, a detailed 
analysis, however, still needs substantial work. Even then seems an
explicit evaluation in closed analytical form for all realistic
potentials prohibitive. Our intention here is to give 
for a highly non-trivial  example a first account about  the
significant role of quantum fluctuations in the QSR.  

To summarize we have studied for two important examples quantum transport over
driven barrier potentials in the strong friction limit. For the case
of a fluctuating bistable potential we explored the impact of quantum
fluctuations on the resonant 
activation phenomenon. In a case of a metastable potential driven
externally by a periodic force we found a sensitive behavior of the
decay rate on the relevant parameters. These notables findings may
stimulate further studies of the quantum Smoluchowski equation in
physics and chemistry as well.

\section*{Acknowledgments}
Valuable discussions with P.\ Pechukas and H.\ Grabert are gratefully
acknowledged. This work was supported by the DFG (Bonn) through SFB276.

\end{multicols}


\begin{thebibliography}{99}
\bibitem{kramers} P.\ H\"anggi, P.\ Talkner, M.\ Borkovec, Rev.\ Mod.\
Phys. {\bf  62}, 251 (1990).
\bibitem{jung} L.\ Gammaitoni, P.\ H\"anggi, P.\ Jung, and F.\ Marchesoni,
Rev.\ Mod.\ Phys.\ {\bf 70}, 223 (1998).
\bibitem{fluc1d} C.R.\ Doering, J.C.\ Gadoua, Phys.\ Rev.\ Lett.\ {\bf
69}, 2318 (1992); P.\ Pechukas, P.\ H\"anggi, {\it ibid}
{\bf 73}, 2772 (1994); P.\ Reimann, T.\ Elston, {\it ibid} {\bf 77},
5328 (1996);R.N.\ Mantegna, B.\ Spagnolo, Phys.\ Rev.\ Lett.\ {\bf 84},
3025 (2000).
\bibitem{astumian} R.D.\ Astumian, Science {\bf 276}, 917 (1997).
\bibitem{weiss} U.\ Weiss, {\it Quantum Dissipative Systems}, (Singapore,
World Scientific, 1999).
\bibitem{redfield} W.H.\ Louisell, {\it Quantum Statistical Properties
of Radiation} (Wiley, New York, 1973).
\bibitem{makri} D.\ E.\ Makarov and N.\ Makri, Chem.\ Phys.\
Lett. {\bf 221}, 482 (1994).
\bibitem{qsr} M.\ Thorwart and P.\ Jung, Phys.\ Rev.\ Lett.\ {\bf 78},
2503 (1997). M.\ Grifoni and P.\ H\"anggi, Phys.\ Rep.\ {\bf 304}, 230
(1998) and references therein;M.\ Thorwart, M.\ Grifoni, and P.\ H\"anggi,
Phys. Rev. Lett. {\bf 85}, 860 (2000).
\bibitem{prl} J.\ Ankerhold, P.\ Pechukas, and H.\ Grabert,
Phys. Rev. Lett. {\bf 87}, 086802 (2001).
\bibitem{remark} As in the classical Smoluchowski limit steep
potentials are excluded from this study.
\bibitem{physica} J.\ Ankerhold and P.\ Pechukas, Physica A {\bf 261},
458 (1998);Phys.\ Rev.\ E {\bf 58}, 6968 (1998).
\bibitem{osci} V.N.\ Smelyanskiy, M.I.\ Dykman, B.\ Golding, Phys.\ Rev.\
Lett.\ {\bf 82}, 3193 (1999).
\bibitem{lehmann} J.\ Lehmann, P.\ Reimann, P.\ H\"anggi,
Phys.\ Rev.\ Lett.\ {\bf 84}, 1639 (2000).
\bibitem{stein} R.S.~Maier and D.L.~Stein, Phys.\ Rev.\ Lett.\ {\bf
86}, 3942 (2001).



\end{thebibliography}
\end{document}